\documentclass[twocolumn,prl,showpacs]{revtex4}
\usepackage{amssymb,amsmath,graphicx}

\begin{document}

\title{External Time-Varying Fields and Electron Coherence}
\author{Jen-Tsung Hsiang}
\email[Jen-Tsung Hsiang: ]{jen-tsung.hsiang@tufts.edu}
\author{L. H. Ford}
\email[L. H. Ford: ]{ford@cosmos.phy.tufts.edu}
\affiliation{Institute of Cosmology, Department of Physics and Astronomy\\
        Tufts University\\
        Medford, Massachusetts, 02155}
\date{\today}

\begin{abstract}
    The effect of time-varying electromagnetic fields on electron
    coherence is investigated. A sinusoidal electromagnetic field
produces a time varying Aharonov-Bohm phase. In a measurement of
the interference pattern which averages over this phase, the
effect is a loss of contrast. This is effectively a form of
decoherence. We calculate the magnitude of this effect for various
electromagnetic field configurations. The result seems to be
sufficiently large to be observable.
\end{abstract}

\maketitle

The well-known Aharonov-Bohm phase~\cite{AB1959} arises when coherent
electrons traverse two distinct paths in the presence of an electromagnetic
field. Let the two paths in spacetime be denoted by $C_1$ and $C_2$.
The phase difference due to the electromagnetic field, the Aharonov-Bohm
phase, is the line integral of the vector potential around the closed
spacetime path $\partial\Omega=C_1-C_2$:
\begin{equation}
    \vartheta=-e\oint_{\partial\Omega} dx_{\mu}A^{\mu}(x)\,.
               \label{E:phase_difference}
\end{equation}
By Stoke's theorem, it can also be expressed as a surface integral
of the field strength tensor over a two dimensional surface $\Omega$
bounded by $\partial\Omega$:
\begin{equation}
    \vartheta=
-\frac{1}{2} e\int_{\Omega} d\sigma_{\mu \nu}\, F^{\mu \nu}(x)\,.
               \label{E:phase_difference2}
\end{equation}
This leads to the remarkable result that the electron interference
pattern is sensitive to shifts in the field strength in regions
from which the electrons are excluded. The reality of the
Aharonov-Bohm effect has been confirmed by numerous experiments,
beginning with the work of Chambers~\cite{rC1960} and continuing
with that of Tonomura and coworkers~\cite{aT1982} using electron
holography.

If the electromagnetic field undergoes fluctuations on a time
scale shorter than the integration time of the experiment, then
the effect is a loss of contrast in the interference pattern. The
role of a fluctuating Aharonov-Bohm phase in decoherence has been
discussed by several
authors~\cite{SAI1990,lF1993,lF1995,lF1997,BP2001,BP2000,MPV2003}.
The amplitude of the interference oscillations is reduced by a
factor of
\begin{equation}
\Upsilon = \left\langle {\rm e}^{i\vartheta}\right\rangle \,,
                                               \label{eq:upsilon}
\end{equation}
where the angular brackets can denote either an ensemble or a time average.
In the case of Gaussian or quantum fluctuations with
$\langle \vartheta \rangle=0$, this factor becomes
\begin{equation}
\Upsilon = {\rm e}^{-\frac{1}{2}\langle \vartheta^2 \rangle} \, .
                                        \label{eq:Gaussian_fluct}
\end{equation}
This form also holds in the case of thermal fluctuations~\cite{BP2001}.

In our treatment, we assume an approximation in which the electrons move
on classical trajectories. More generally, the electrons are in wavepacket
states. However, under many circumstances, the sizes of the wavepackets
can be small compared to the path separation, so the classical path
approximation is good. Wavepacket sizes which have been realized in
experiments~\cite{NH93} can be less than $1\, {\rm \mu m}$, which is
one to two orders of magnitude smaller than the other length scales 
characterizing the paths. A more detailed discussion of the effects of finite
wavepacket size was given in Ref.~\cite{lF1997}. 

The purpose of the present paper is to discuss a particularly simple
version of this type of decoherence produced by a classical, sinusoidal
electromagnetic field. If the period of oscillation of the field is
short compared to the time scale over which the interference pattern can
be measured, then a time average must be taken in Eq.~(\ref{eq:upsilon}),
with a resulting loss of contrast.

We consider the case of a linearly polarized, monochromatic electromagnetic
wave of frequency $\omega$ which propagates in a direction perpendicular
to the plane containing the electron beams. Let the wave be polarized
in the $z$-direction and propagate in the $y$-direction, with the
plane of the electron paths being the $x$-$z$ plane. For a path confined
to this plane, we have
\begin{equation}
\frac{1}{2} \,d\sigma_{\mu \nu}\, F^{\mu \nu} = dt\, dx \, F^{tx} +
dt\, dz \, F^{tz} + dx\, dz \, F^{xz} \,.
\end{equation}
In the present case, where $E^x = B^y =0$, Eq.~(\ref{E:phase_difference2})
becomes
\begin{equation}
    \vartheta= e \int dt\,dz \,E^z  \,.
               \label{E:phase_difference3}
\end{equation}
Let the $z$-component of the electric
field take the form
\begin{equation}
    E^z(x^{\mu})=\mathbb{E}(x,y,z)\cos(k\,y-\omega\,t), \label{eq:Efield}
\end{equation}
where the real modulated amplitude $\mathbb{E}(x,y,z)$ is assumed to be
a slowly varying function of $y$, compared with the sinusoidal
oscillation.
We can write
\begin{equation}\label{E:vartheta}
    \vartheta(t_0) = e\!\int_{\Omega}dt\,dz\,
\mathbb{E}(x,y,z)\cos(k\,y-\omega\,t-\omega\,t_0)\, ,
\end{equation}
where $t_0$ is the electron emission time. More precisely, it is the
time at which the center of a localized wavepacket is emitted. 
If the measuring
process takes a sufficiently long time compared with the electron
flight time, we will observe a result which is averaged over
$t_0$. Therefore, let $t_0$ be a random variable and take the time
average over that variable. That is, for a function $f$ of a
random time variable $\xi$, the time average is defined by
\begin{equation}
    \bigl<f(\xi)\bigr>\equiv\lim_{\Xi\rightarrow\infty}
\frac{1}{2\,\Xi}\int_{-\Xi}^{+\Xi}d\xi\,f(\xi).
\end{equation}
However, before taking the time average, we will rewrite
Eq.~\eqref{E:vartheta} as
\begin{equation}\label{E:vartheta_1}
    \vartheta=\mathbb{A}\cos(\omega\,t_0)+\mathbb{B}\sin(\omega\,t_0),
\end{equation}
where
\begin{align}
    &\mathbb{A}=e\!\int_{\Omega}dt\,dz\,\mathbb{E}(x,y,z)\cos(k\,y-\omega\,t),\\
    &\mathbb{B}=e\!\int_{\Omega}dt\,dz\,\mathbb{E}(x,y,z)\sin(k\,y-\omega\,t),
\end{align}
and we have the average of the time-varying phase factor given by,
\begin{align}
  \Upsilon = \bigl<e^{i\vartheta}\bigr>  &=\lim_{\Xi\rightarrow\infty}\frac{1}{2\,\Xi}\!
\int_{-\Xi}^{+\Xi}\!dt_0\;e^{i\,\bigl[\mathbb{A}\cos(\omega\,t_0)+
\mathbb{B}\sin(\omega\,t_0)\bigr]},\notag\\
    &=J_0\bigl(\left|\,\mathbb{C}\,\right|\bigr), \label{eq:Upsilon}
\end{align}
where $J_0$ is a Bessel function and
\begin{align}\label{E:const_C}
    \mathbb{C}  &=\mathbb{A}+i\,\mathbb{B}\notag\\
                &=e\!\int_{\Omega}dt\,dz\,\mathbb{E}(x,y,z)\,e^{i(k\,y-\omega\,t)}.
\end{align}

Note that in the limit that $|\mathbb{C}| \ll 1$, we can Taylor expand the
Bessel function $J_0$ and write
\begin{equation}
\Upsilon \approx 1 - \frac{1}{4} |\mathbb{C}|^2 + \frac{1}{64} |\mathbb{C}|^4
+ \cdots \,.
\end{equation}
This agrees through order $|\mathbb{C}|^2$ with the result that would be
obtained from Eq.~(\ref{eq:Gaussian_fluct}) for Gaussian fluctuations,
as $\langle \vartheta^2 \rangle =\frac{1}{2} |\mathbb{C}|^2$.

As the strength of the applied field increases, the contrast factor
$\Upsilon$ will monotonically decrease until the first zero of $J_0$
at $|\mathbb{C}| = 2.405$ is reached. Beyond that point, the contrast will
begin to increase and then undergo damped oscillations. This behavior
is quite different from that produced by Gaussian fluctuations,
Eq.~(\ref{eq:Gaussian_fluct}).

Now we study the possible effect on the electron interference if
we shine a non-localized beam over the electron paths. Because the
plane wave extends to infinity in the transverse direction, it is
inevitable that the electron will have direct interaction with the
electromagnetic fields, However, it will be shown later that
 the direct interaction with the electromagnetic fields
is extremely small, so it can be ignored. Some years ago, Dawson
and Fried~\cite{DF1967} discussed the effect of a laser beam on
coherent electrons. However, these authors were concerned with a
change in phase, rather than the loss of contrast with which we
are concerned.

\begin{figure}
\centering
    \includegraphics{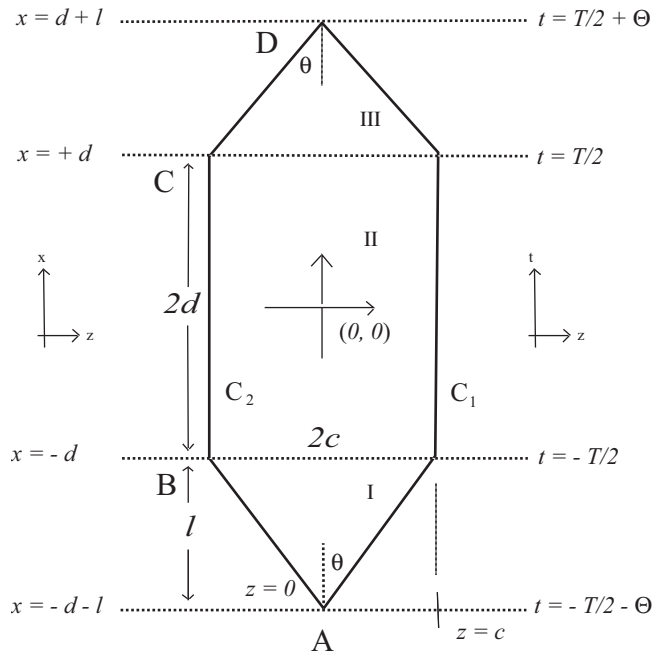}
    \caption{The two possible electron paths, $C_1$ and $C_2$ are
illustrated. The electrons start at point $A$ and end at point $D$
after traversing a path which is approximated by three straight
line segments. Here $\Theta$ is the time required for the first and last
segments, and $T$ is the time required for the middle segment.}
\label{Fi:fig_1}
\end{figure}

Assume that the transverse plane wave of amplitude $\mathcal{E}_0$
propagates along the $y$ axis and is polarized in the $z$
direction. The electron paths lie on the $y=0$ plane and are illustrated
in Fig.~\ref{Fi:fig_1}. The quantity $\mathbb{C}$ is then given by
\begin{equation}
    \mathbb{C}=4\,e\,\mathcal{E}_0\biggl(\frac{2c}{\omega^2\Theta}\biggr)
\sin\biggl[\frac{\omega\,\Theta}{2}\biggr]\sin\biggl[\frac{1}{2}\,
\omega\left(T+\Theta\right)\biggr].
\end{equation}
Here $2c$ is the maximum separation between the elctron paths. Experimentally
attainable separations are of the order of $100\, \mu
m$~\cite{Hasselbach}.

The quantity
$\left|\,\mathbb{C}\,\right|^2$  is written as
\begin{align}\label{E:C2}
    \left|\,\mathbb{C}\,\right|^2
&=16\,e^2\mathcal{E}_0^2\biggl(\frac{2c}{\omega^2\Theta}\biggr)^2\sin^2
\biggl[\frac{\omega\,\Theta}{2}\biggr]\sin^2\biggl[\frac{1}{2}\,
\omega\left(T+\Theta\right)\biggr]\notag\\
 &\approx\frac{32\pi}{137}\,\rho\,\biggl(\frac{2c}{\omega^2\Theta}\biggr)^2\, ,
\end{align}
where the squares of the sine functions have been replaced by their average
value of $1/2$ and the averaged energy density $\rho$ is given by
\begin{equation}
    \rho=\frac{1}{2}\,\mathcal{E}_0^2.
\end{equation}
We use Lorentz-Heaviside units with $\hbar$ and the speed of light set equal to
unity. Thus, $\rho$ is also the energy flux in the electromagnetic wave.
Note that $\Theta = s/v$, where $v$ is the electron's speed and
$s=\sqrt{c^2+l^2}$ is the length of the first and third segments of the paths.
If the electron's speed is nonrelativistic, we can write
\begin{equation}
\left|\,\mathbb{C}\,\right|^2 = \left(\frac{E_k}{5\,{\rm keV}}\right)
\left(\frac{\rho}{1\,{\rm W/cm}^2}\right) \left(\frac{2c}{s}\right)^2
\left(\frac{\lambda}{100\,\mu {\rm m}}\right)^4 \, ,
\end{equation}
where $E_k$ is the electron kinetic energy and $\lambda$ is the
wavelength of the electromagnetic wave. Thus it seems plausible
that one could arrange to have $\left|\,\mathbb{C}\,\right|^2$
large enough to produce experimentally observable effects. 

There are some comments on this calculation: First, we assumed electron
paths with sharp corners for simplicity. If one were to round out the
corners slightly to make more realistic paths, the result need not change
significantly. This is because we are integrating a regular integrand which
varies on a time scale of the order of $1/\omega$. If the actual time scale
for the electron to change direction is small compared to this time, then
our piecewise trajectory is a good approximation. Note that here we are
discussing the change in contrast due to the applied field. Sharp corners
will tend to cause emission of photons, which in turn lead to decoherence
even in the absence of an applied field.
 A second comment is that the
contributions of each of the three regions, I, II and III, in
Fig.~\ref{Fi:fig_1} is large compared to the final result for
$\mathbb{C}$ by a factor of the order of $\Theta\, \omega$.
However, the leading terms cancel when the three contributions are
summed, leading to Eq.~(\ref{E:C2}). Finally, we have assumed that
the electron paths are localized, whereas in an actual experiment
the classical trajectories will be replaced by bundles of finite
thickness. What is required here is that the electron beams be
localized in the $y$-direction on a scale small compared to the
wavelength of the electromagnetic field.

Since the electron passes through the region where the
electromagnetic fields are non-zero, it has a direct interaction
with the fields. Due to the fact that the electron is in
non-relativistic motion, in the low energy limit, only Thomson
scattering is considered.
Let $n$ be the mean number density of photons, which can
approximately be expressed in terms the electromagnetic energy
density $\rho$ and the angular frequency $\omega$ as
\begin{equation}
    n\simeq\frac{\rho}{\omega},
\end{equation}
for very large $n$. As a result, the mean free path $l_{\rm mfp}$ of
the Thomson scattering is given by
\begin{align}
    l_{\rm mfp} &=\frac{1}{n\,\sigma_T}=\frac{\omega}{\sigma_T\,\rho}\\
            &=9\times10^{13}\textit{m}\,
\Bigl(\frac{\rho}{{\rm W/cm}^2}\Bigr)^{-1}\Bigl(\frac{\lambda} {\mu
{\rm m}}\Bigr)^{-1}\,,
\end{align}
where $\sigma_T$ is the Thomson cross section.
We can see that it is possible to have an incident flux
which is large enough to produce observable decoherence but for which any
effect from the electron-photon scattering may be ignored. 
That is, loss of phase coherence due to direct electron-photon
scattering arises from the random accumulated electron wavefunction phase
shifts from one or more such scattering events. However, in many realistic 
situations, the probability of even one such event per electron is close
to zero.

The above analysis shows that the change of contrast is really due to
a variant of the Aharonov-Bohm effect, the averaging over the time-dependent
Aharonov-Bohm phase created by fields in the interior of the electron path.
It is not due to direct scattering between electrons and photons.
Nonetheless, it is also of interest to consider a configuration where the
applied electromagnetic field is localized in a region between the
electron paths. An example is a Gaussian
beam. Let the electric field in the plane of the paths be given by
\begin{equation}
    E^z({\bf \rho})= \mathcal{E}_0
\,\exp\left(-\frac{{\boldsymbol{\rho}}^2}{\sigma^2}\right)\,
                  \cos(\omega\,t) \, ,
                                  \label{eq:Gauss_field}
\end{equation}
where $\boldsymbol{\rho}$ is the radius vector in the plane and $\sigma$
is the effective width of the beam in this plane. This form is a
good approximation to the electric field of a linearly polarized
laser beam. Suppose that this beam is normally incident upon the
electron paths illustrated in Fig.~\ref{Fi:fig_1}, with the center
of the beam being at the origin in this figure. A calculation
which will be presented in detail in Ref.~\cite{HF2004} leads to
the result, for the case that $\sigma \alt 2 c$ and $\sigma \alt 2 d$,
\begin{equation}
\mathbb{C} \approx - \frac{8\sqrt{\pi} e \mathcal{E}_0 d^2}{\omega^2 T \sigma}
\, (1-\cos\theta)\, \cos\left(\frac{\omega T}{2}\right) \,
\exp\left(-\frac{d^2}{\sigma^2}\right) \,.
\end{equation}
The crucial feature of this result is the factor of $\exp(-d^2/\sigma^2)$,
which is extremely small in the limit of a highly localized beam,
$\sigma \ll d$.

To summarize, in this paper we have investigated the effects of a rapidly
varying Aharonov-Bohm phase upon an electron interference pattern. If the
time scale for the variation is short compared to the time during which
the pattern is measured, then averaging over the phase variations leads
to a loss of contrast. This is a form of decoherence. In principle, the
lost contrast could be restored if one were able to select only those electrons
which start at a fixed point in the cycle of an oscillatory Aharonov-Bohm phase.
The form of decoherence studied here is an example of zero temperature 
decoherence. Other forms of zero temperature decoherence, which do not rely
upon thermal effects, have been discussed in Refs.~\cite{Sinha97,WMJ98,WG01}.

We have calculated the size of the decoherence effect produced by
a monochromatic, linearly polarized electromagnetic field. The
result seems to be large enough to be observable. We primarily treated
the case of a non-localized plane wave. In this case, although the
electromagnetic field is nonzero at the location of the electrons,
we argued that one can have an observable loss of contrast even
when the probability of an electron scattering from a photon is
extremely small.  A unique signature of the
decoherence produced by sinusoidal fields is that the interference
pattern can disappear and then reappear as the field strength is
increased.

\begin{acknowledgments}
We would like to thank Ken Olum for valuable discussion. 
 This work was supported in part by the National
Science Foundation under Grant PHY-0244898.
\end{acknowledgments}

\end{document}